%% file: cancellation_iscas.tex
\documentclass[conference]{IEEEtran} % Uses 10 pt
\IEEEoverridecommandlockouts

\usepackage[T1]{fontenc}
\usepackage{textcomp}
\usepackage[utf8]{inputenc}

\usepackage{stfloats}
\usepackage[normalem]{ulem}

\usepackage{amsthm}
\usepackage{mathtools}
\usepackage{amssymb}
\usepackage{amsfonts}
\usepackage{xfrac}
\usepackage{bm} % Note that you may not always want to use \bm <https://tex.stackexchange.com/questions/3238/bm-package-versus-boldsymbol>

\newcommand{\gd}{g_\text{d}}
\newcommand{\gi}{g_\text{i}}
\newcommand{\gs}{g_\text{s}}

\newcommand{\hdelay}{h_{\text{delay}}}
\newcommand{\htxzero}{h_{\text{tx,0}}}
\newcommand{\hchzero}{h_{\text{ch,0}}}
\newcommand{\hrx}{h_{\text{rx}}}
\newcommand{\hfilt}{h_{\text{filt}}}
\newcommand{\htxone}{h_{\text{tx,1}}}
\newcommand{\hchone}{h_{\text{ch,1}}}

\newcommand{\htxi}{h_{\text{tx,i}}}
\newcommand{\hchi}{h_{\text{ch,i}}}

\newcommand{\rzero}{r_0}
\newcommand{\rone}{r_1}
\newcommand{\rfilt}{r_{\text{filt}}}
\newcommand{\rcanc}{r_{\text{canc}}}

\newcommand{\boldhfilt}{\mathbf{h}_{\text{filt}}}
\newcommand{\boldrone}{\mathbf{r}_1}

\newcommand{\errorterm}{e}

\usepackage{lipsum}

\usepackage[usenames,dvipsnames]{xcolor}

\usepackage{tikz}
\usepackage{circuitikz}
\usepackage{pgfplots}
\pgfplotsset{compat=1.15}
\usepgfplotslibrary{units}
\usepgfplotslibrary{groupplots}
\usepgfplotslibrary{colormaps}
\usepackage{pgfplotstable}

\usetikzlibrary{calc}
\usetikzlibrary{shapes}
\usetikzlibrary{shapes.misc}
\usetikzlibrary{matrix}
\usetikzlibrary{arrows}
\usetikzlibrary{decorations.markings}
\usetikzlibrary{spy}
\usetikzlibrary{decorations.pathmorphing}
\usetikzlibrary{decorations.markings}
\usetikzlibrary{positioning}
\usetikzlibrary{fit}
\usetikzlibrary{petri}
\usetikzlibrary{chains}
\usetikzlibrary{intersections}
\usetikzlibrary{plotmarks}
\usetikzlibrary{backgrounds}
\tikzset{>=latex}

\usepackage{multicol}
\usepackage{graphicx}
\usepackage{float}

\def\BibTeX{{\rm B\kern-.05em{\sc i\kern-.025em b}\kern-.08em
    T\kern-.1667em\lower.7ex\hbox{E}\kern-.125emX}}

\usepgfplotslibrary{statistics}

\usepackage{subcaption}
\usepackage{caption}

\usepackage{array}
\usepackage{csvsimple}
\usepackage{booktabs}
\usepackage{ifthen}
\usepackage{multirow}
\usepackage{makecell}
\usepackage{boldline}
\usepackage{threeparttable}
\usepackage{pbox}
\usepackage{balance}

\usepackage{url}
\usepackage{verbatim}

\usepackage{algorithm}% http://ctan.org/pkg/algorithms
\usepackage{algpseudocode}% http://ctan.org/pkg/algorithmicx

\usepackage[%
  square,        % for square brackets
  comma,         % use commas as separators
  numbers,       % for numerical citations;
  sort&compress % as sort but in addition multiple numerical citations
]{natbib}

\usepackage[super]{nth}
\usepackage[detect-weight=true, per-mode=symbol,group-separator={,},group-minimum-digits={3}]{siunitx}[=v2]
\DeclareSIUnit{\belmilliwatt}{Bm}
\DeclareSIUnit{\dBm}{\deci\belmilliwatt}
\DeclareSIUnit{\bpm}{bpm}
\DeclareSIUnit{\dBc}{dBc}
\DeclareSIUnit{\dBi}{dBi}

\usepackage[shortcuts]{extdash} % https://tex.stackexchange.com/questions/103608/how-to-force-latex-not-to-break\=/the-line-after-a-hyphen

\usepackage{glossaries}
\glsdisablehyper % Ensure that hyperlinks are not formed if hyperref is loaded

\newacronym{svd}{SVD}{singular value decomposition}
\newacronym{ls}{LS}{least squares}
\newacronym{lms}{LMS}{least mean squares}

\newacronym{cnc}{CNC}{computer numerical control}
\newacronym{usrp}{USRP}{Universal Software Radio Peripheral}
\newacronym{udp}{UDP}{user datagram protocol}
\newacronym{fpga}{FPGA}{field-programmable gate array}
\newacronym{ni}{NI}{National Instruments}

\newacronym{agc}{AGC}{automatic gain control}
\newacronym{rf}{RF}{radio frequency}
\newacronym{tx}{Tx}{transmit}
\newacronym{rx}{Rx}{receive}
\newacronym{cdd}{CDD}{cyclic delay diversity}
\newacronym{adc}{ADC}{analog-to-digital converter}
\newacronym{dac}{DAC}{digital-to-analog converter}
\newacronym{mac}{MAC}{medium-access control}
\newacronym{phy}{PHY}{physical}
\newacronym{los}{LOS}{line-of-sight}
\newacronym{nlos}{NLOS}{non-line-of-sight}
\newacronym{si}{SI}{self-interference}
\newacronym{sic}{SIC}{self-interference cancellation}
\newacronym{ansic}{AnSIC}{analog self-interference cancellation}
\newacronym{fd}{FD}{full-duplex}
\newacronym{pa}{PA}{power amplifier}
\newacronym{cp}{CP}{cyclic prefix}
\newacronym{mimo}{MIMO}{multiple-input multiple-output}
\newacronym{ofdm}{OFDM}{orthogonal frequency-division multiplexing}

\newacronym{iq}{I/Q}{in-phase and quadrature}
\newacronym{csi}{CSI}{channel state information}
\newacronym{cir}{CIR}{channel impulse response}
\newacronym{cfr}{CFR}{channel frequency response}
\newacronym{dft}{DFT}{discrete Fourier transform}
\newacronym{idft}{IDFT}{inverse discrete Fourier transform}
\newacronym{fft}{FFT}{fast Fourier transform}
\newacronym{pca}{PCA}{principal component analysis}
\newacronym{ica}{ICA}{independent component analysis}
\newacronym{bss}{BSS}{blind source separation}
\newacronym{music}{MUSIC}{multiple signal classification}
\newacronym{trrs}{TRRS}{time-reversal resonating strength}
\newacronym{lte}{LTE}{TODO}
\newacronym{nr}{NR}{TODO}

\newacronym{uwb}{UWB}{ultra-wideband}
\newacronym{fmcw}{FMCW}{frequency modulated continuous wave}
\newacronym{isac}{ISAC}{integrated sensing and communication}
\newacronym{sdr}{SDR}{software-defined radio}
\newacronym{vm}{VM}{vector modulator}
\newacronym{pcb}{PCB}{printed circuit board}
\newacronym{ebd}{EBD}{electrical balance duplexer}

\newacronym{cfo}{CFO}{carrier frequency offset}
\newacronym{sfo}{SFO}{sampling frequency offset}
\newacronym{sto}{STO}{sampling time offset}
\newacronym{iip3}{IIP3}{third-order input intercept point}
\newacronym{cpe}{CPE}{common phase error}

\newacronym{aoa}{AoA}{angle of arrival}
\newacronym{boi}{BoI}{band-of-interest}
\newacronym{bnr}{BNR}{breathing-to-noise ratio}
\newacronym{bpm}{bpm}{breaths per minute}
\newacronym{psd}{PSD}{power spectral density}
\newacronym{snr}{SNR}{signal-to-noise ratio}
\newacronym{mse}{MSE}{mean square error}

\newacronym{lstf}{L-LTF}{legacy short training field}
\newacronym{lltf}{L-LTF}{legacy long training field}
\newacronym{htstf}{HT-STF}{high throughput short training field}
\newacronym{htltf}{HT-LTF}{high throughput long training field}
\newacronym{lsig}{L-SIG}{legacy signal field}
\newacronym{htsig}{HT-SIG}{high throughput signal field}
\newacronym{ht}{HT}{high throughput}

\newacronym{cd}{CD}{channel decomposition}
\newacronym{pc}{PC}{phase calibration}
\newacronym{midarr}{MIDARR}{Monostatic Interference Decomposition and Respiration Recovery}

\input{macros}

\input{colorscheme}

\newcommand{\Colorcoffpoffdoff}{thsViolet4}

\newcommand{\Colorcoffpondon}{thsOrange5}
\newcommand{\Colorconpoffdoffgoff}{thsRed3}

\newcommand{\Colorconpoffdoffgon}{thsBrown4}

\newcommand{\andreas}[1]{\textcolor{Green}{ANDREAS: #1}}
\newcommand{\sitian}[1]{\textcolor{blue}{SITIAN: #1}}
\newcommand{\alex}[1]{\textcolor{purple}{ALEX: #1}}
\newcommand{\andy}[1]{\textcolor{orange}{ANDY: #1}}
\newcommand{\question}[1]{\textcolor{BlueViolet}{\textbf{QUESTION: #1}}}
\newcommand{\todo}[1]{\textcolor{red}{\textbf{TODO: #1}}}

\newcommand{\mycomment}[1]{}

\renewcommand{\andreas}[1]{}
\renewcommand{\sitian}[1]{}
\renewcommand{\alex}[1]{}
\renewcommand{\andy}[1]{}
\renewcommand{\question}[1]{}
\renewcommand{\todo}[1]{}
\newcommand{\figurewidthninetyfive}{240pt}  % 95%  of \columnwidth
\newcommand{\figurewidtheighty}{200pt}      % 80%  of \columnwidth
\newcommand{\figurewidthsixtyfive}{157pt}   % 65%  of \columnwidth
\newcommand{\figurewidthsixty}{150pt}       % 60%  of \columnwidth
\newcommand{\figurewidthfiftytwo}{130pt}    % ~52% of \columnwidth
\newcommand{\figurewidthfifty}{125pt}       % ~50% of \columnwidth
\newcommand{\figureheightsixty}{100pt}      % 60%  of \figurewidthsixty
\newcommand{\figureheightfiftytwo}{82pt}    % ~52% of \figurewidthfiftytwo
\newcommand{\figureheightfifty}{80pt}       % ~50% of \figurewidthfifty
\newcommand{\figuretextsize}{\small}

\newcommand{\subfigurevstackvspace}{0.1cm} % Vertical space between subfigures in a vertical stack
\newcommand{\subfigurecaptionspace}{0.4cm} % How much to move small captions for subfigures
\newcommand{\subfigurecaptionspacesmall}{0.2cm}
\newcommand{\groupplotsep}{4pt} % Separation between group plots

\pgfplotsset{
  pgfstyledefault/.style = {
    minor x tick num=1,
    xtick pos=left,
    ytick pos=left,
    every x tick/.style={color=black, thick},
    every y tick/.style={color=black, thick},
    axis line style = thick,
    tick align=outside,
    xlabel near ticks,
    ylabel near ticks,
    axis on top,
    enlarge x limits=false,
    enlarge y limits=true,
  }
}

\pgfplotsset{
  pgfstyleonedspectrum/.style = {
    pgfstyledefault,
    width=\figurewidthninetyfive,
    height=\figureheightfiftytwo,
    xlabel={Frequency [bpm]},
    xmin=0.1, xmax=39.9, % This ensures that the 0 and 40 are not plotted, to align the figures properly
    xtick distance=5,
    ylabel={Power [dB]},
    enlargelimits=false,
  }
}

\pgfplotsset{
  pgfstyleheatmap/.style = {
    pgfstyledefault,
    enlargelimits=false,
    colormap/jet,
    colorbar,
    disabledatascaling,
    colorbar/width=3.0mm,
  }
}

\pgfplotsset{
  pgfstyletwodspectrum/.style = {
    pgfstyleheatmap,
    width=\figurewidtheighty,
    height=\figureheightfiftytwo,
    xlabel={Frequency [bpm]},
    xmin=0.1, xmax=39.9, % This ensures that the 0 and 40 are not plotted, to align the figures properly
    xtick distance=5,
    ylabel={Delay [ns]},
    point meta min=-40,
    point meta max=0,
    colorbar style={
      yticklabel style={text width=width("$-20$"), align=right} % Compensate for the negative sign
    },
  }
}

\pgfplotsset{
  pgfstyletwodspectrumnormtop/.style = {
    pgfstyletwodspectrum,
    width=\figurewidthninetyfive,
    height=\figureheightfiftytwo,
    xlabel={},
    xtick=\empty,
    colorbar horizontal,
    colorbar style={
      at={(0.5,1.1)},
      anchor=south,
      xticklabel pos=upper,
      width=0.95*\pgfkeysvalueof{/pgfplots/parent axis width},
    },
  }
}

\pgfplotsset{
  pgfstyletwodspectrumnormmiddle/.style = {
    pgfstyletwodspectrum,
    width=\figurewidthninetyfive,
    height=\figureheightfiftytwo,
    xlabel={},
    xtick=\empty,
    colorbar=false,
    colorbar style={},
  }
}

\pgfplotsset{
  pgfstyletwodspectrumnormbottom/.style = {
    pgfstyletwodspectrum,
    width=\figurewidthninetyfive,
    height=\figureheightfiftytwo,
    colorbar=false,
    colorbar style={},
  }
}

\begin{document}

\title{An SDR-Based Monostatic Wi-Fi System with Analog Self-Interference Cancellation for Sensing}

\author{\IEEEauthorblockN{
Andreas Toftegaard Kristensen\IEEEauthorrefmark{1},
Alexios Balatsoukas-Stimming\IEEEauthorrefmark{2},
and Andreas Burg\IEEEauthorrefmark{1}
}
\IEEEauthorblockA{%
\IEEEauthorrefmark{1}Telecommunication Circuits Laboratory, \'{E}cole polytechnique f\'{e}d\'{e}rale de Lausanne, Switzerland\\
\IEEEauthorrefmark{2}Eindhoven University of Technology, The Netherlands\\
}}

\maketitle

\begin{abstract}
Wireless sensing offers an alternative to wearables for contactless monitoring of human activity and vital signs.
However, most existing systems use bistatic setups, which suffer from phase imperfections due to unsynchronized clocks.
Monostatic systems overcome this issue, but are hindered by strong self-interference (SI) that require effective cancellation.
We present a monostatic Wi-Fi sensing system that uses an auxiliary transmit RF chain to achieve SI cancellation levels of 40 dB, comparable to existing solutions with custom cancellation hardware.
We demonstrate that the cancellation filter weights, fine-tuned using least-mean squares, can be directly repurposed for target sensing.
Moreover, we achieve stable SI cancellation over 30 minutes in an office environment without fine-tuning, enabling traditional vital sign monitoring using channel estimates derived from baseband samples without the adaptation of the cancellation affecting the sensing channel -- a significant limitation in prior work.
Experimental results confirm the detection of small, slow-moving targets, representative for breathing chest movements, at distances up to 10 meters in non-line-of-sight conditions.
\end{abstract}

%%%%%%%%%%%%%%%%%%%%%%%%%%%%%%%%%%%%%%%%%%%%%%%%%%%%%%%%%%%%%%%%%%%%%%%%%%%%%%%
\section{Introduction}\label{sec:introduction}

% Motivation: Trend in detecting vital signs and wearables limitations
Real-time, continuous monitoring of human activity and vital signs, particularly heart and respiratory rates, is crucial for enhancing healthcare analytics and early detection of health issues~\cite{gambhirAchievingPrecisionHealth2018, khanMonitoringVitalSigns2016, ebellPredictingPneumoniaAdults2007, fox2007resting}.
However, wearable technologies face limitations such as user discomfort and non-compliance~\cite{ometovSurveyWearableTechnology2021, jeong2017smartwatch}.
An ideal monitoring system would enable contactless, continuous data collection with minimal discomfort for the user.
%
% Related work: Opportunistic sensing and ISAC systems
Recognizing these challenges, researchers have focused on developing wireless contact-free monitoring systems using common technologies such as Wi-Fi.
Most Wi-Fi sensing systems extract channel information from commodity devices, avoiding specialized hardware~\cite{halperinToolReleaseGathering2011, xiePrecisePowerDelay2015, gringoliFreeYourCSI2019, jiangEliminatingBarriersDemystifying2021, gringoliAXCSIEnablingCSI2022}.
However, these systems operate in bistatic mode, with physically separated transmitters and receivers, which face challenges due to unsynchronized transmitter and receiver clocks, causing imperfections such as \gls{cfo}, \gls{sfo}, and \gls{sto} that severely impair sensing performance.
While calibration methods can partially mitigate these issues~\cite{zengFarSensePushingRange2019}, they also lead to information loss and potential distortion of the sensing signal.
In contrast, monostatic setups avoid the synchronization issues, but face other challenges like \gls{si}, which necessitate \gls{sic} strategies similar to those in \gls{fd} radio systems~\cite{duarteFullduplexWirelessCommunications2010}.
Without proper handling, the strong \gls{si} can mask targets, cause receiver non-linearities, and lead to \gls{adc} saturation as the \gls{si} is orders of magnitude stronger than the sensing signal~\cite{korpi2017full, barnetoFullDuplexOFDMRadar2019}.
%
% Related work: Monostatic sensing and cancellation
To mitigate \gls{si} for both sensing and communication, various \gls{sic} techniques have been explored.
Direct \gls{rf} cancellation methods rely on analog electronics that tap into the \gls{rf} signal from the \gls{tx} chain to train and drive a canceller.
By directly accessing the output of components like \glspl{pa}, direct \gls{rf} cancellers avoid explicitly modeling potential non-idealities, but complex custom cancellation hardware is still needed to adapt the cancellation signal~\cite{korpi2017full}.
A simpler approach in terms of the hardware is to use an auxiliary \gls{tx} path, where the \gls{sic} signal is digitally generated using a secondary \gls{tx} port for \gls{rf} cancellation~\cite{laughlinWidelyTunableFull2015, kiayaniActiveRFCancellation2018, kiayaniAdaptiveNonlinearRF2018}.
However, as this auxiliary \gls{tx} path does not have access to the \gls{rf} \gls{si} signal, it may have to explicitly model non-idealities, e.g., from the \gls{pa} to achieve a sufficient level of \gls{sic}.
While this may seemingly put auxiliary \gls{tx} cancellation systems at a disadvantage compared to direct \gls{rf} cancellation systems, which are also more common for sensing~\cite{barnetoFullDuplexOFDMRadar2019, hassaniInBandFullDuplexRadarCommunication2021, hassaniJointInBandFullDuplex2022, chenISACoTIntegratingSensing2022, chenISACFiEnablingFullfledged2024}, we demonstrate in this work that the auxiliary approach can provide sufficient \gls{sic}.
%
% Multipath
Another consideration for cancellation systems is the calibration and updating/tracking of the cancellation parameters.
If the canceller is frequently fine-tuned to track variations in the system hardware, \gls{sic} remains high, but the target is also suppressed~\cite{barnetoFullDuplexOFDMRadar2019, hassaniInBandFullDuplexRadarCommunication2021, hassaniJointInBandFullDuplex2022, chenISACoTIntegratingSensing2022, chenISACFiEnablingFullfledged2024}.
Proposed mitigation approaches include fitting the canceller to the near environment (a few meters) and only sensing distant targets~\cite{barnetoFullDuplexOFDMRadar2019}, performing Doppler-based sensing for large, fast-moving objects between fine-tuning periods~\cite{hassaniInBandFullDuplexRadarCommunication2021, hassaniJointInBandFullDuplex2022}, or only training the canceller on the direct path \gls{si} using an antenna switch to remove the multipath~\cite{chenISACoTIntegratingSensing2022, chenISACFiEnablingFullfledged2024}.

\subsubsection*{Contributions}
%\gls{ansic}
We present a fully \gls{sdr}-based monostatic Wi-Fi setup with an auxiliary transmit \gls{rf} chain for \gls{sic}, significantly simplifying the hardware requirements compared to previous works that focus on custom analog electronics for \gls{sic}.
We demonstrate that our \gls{ansic} setup achieves \gls{sic} comparable to similar setups and that our cancellation is stable over 30 minutes without fine-tuning the \gls{ansic} filter weights.
Based on this setup, we demonstrate the re-purposing of the \gls{ansic} filter weights for sensing, enabling sensing information recovery at the same time as continuously fine-turning the \gls{ansic} filter weights, which to the best of our knowledge has not been shown before.
Finally, we demonstrate the ability to sense a small slow-moving target at a distance of \SI{10}{\meter} in \gls{nlos} using the residual signal when keeping the filter weights fixed over an extended time period.

\begin{table}[t]
  \centering
  \caption{AnSIC and openwifi FPGA area consumption.}
  \label{tab:fpga_area}
  \begin{tabular}{l|rrrr}
    \toprule
    ~                 & \textbf{LUT} & \textbf{FF} & \textbf{BRAM} & \textbf{DSP} \\
    \gls{ansic}       & \num{8495}        & \num{9983}       & \num{0}              & \num{80} \\
%    openwifi & \num{44753}       & \num{60429}      & \num{126}            & \num{187} \\
    openwifi          & \num{36258}       & \num{50446}      & \num{126}            & \num{107} \\
%    \midrule
%    Total & \num{44753}       & \num{60429}      & \num{126}            & \num{187} \\
    \bottomrule
  \end{tabular}
\end{table}

%%%%%%%%%%%%%%%%%%%%%%%%%%%%%%%%%%%%%%%%%%%%%%%%%%%%%%%%%%%%%%%%%%%%%%%%%%%%%%%
\section{Background}\label{sec:background}

In this section, we first describe the wireless sensing model and then we explain how our \gls{ansic} filter is trained.

%==============================================================================
\subsection{Wi-Fi Sensing}\label{sec:background_wifi_sensing}

For Wi-Fi sensing, while the \gls{cfr} is commonly employed, the \gls{cir} often proves superior as it concentrates sensing information into fewer delay bins, thus reducing feature dimensionality~\cite{liComplexBeatBreathingRate2021, kristensenMonostaticMultiTargetWiFiBased2024}.
The \gls{cir} can be expressed as
\begin{equation}\label{eq:model_cir}
  g(t, \tau) = \sum_{p=1}^{P} \alpha_p(t) \rho(\tau - \tau_p(t)) e^{-j 2\pi f_c \tau_p(t)} \, ,
\end{equation}
where $\rho(\tau) = \frac{\sin(\pi B \tau)}{\pi B \tau}$ is the pulse shape function, $P$ is the number of multipath components, $\alpha_p$ is the coefficient of path $p$, $\tau_p(t)$ is the delay of path $p$, $B$ is the signal bandwidth, and $f_c$ is the carrier frequency.
The \gls{cir} obtained from a wireless device is a sampled \gls{cir} vector $\bmg$, with entries $g[n] = g\left(\frac{n}{B}\right)$.

For a monostatic system, the \gls{cir} can be decomposed into
\begin{equation}\label{eq:model_contributors}
  g(t, \tau) = \gi(t, \tau) + \gs(t, \tau) + \gd(t, \tau) \, ,
\end{equation}
where $\gi(t, \tau)$, $\gs(t, \tau)$, and $\gd(t, \tau)$ represent the channels originating from the \gls{si}, static scatterers, and dynamic scatterers, respectively.
Note that for a more compact notation, we omit the dependency on time $t$ and delay $\tau$ in the following.

%==============================================================================
\subsection{Self-Interference and \gls{ansic} fitting}\label{sec:background_self_interference}

The auxiliary \gls{tx} path model for \gls{ansic} is composed of two paths, with an antenna path for one \gls{dac} and a cancellation path for a second \gls{dac}.
In the antenna path, baseband \gls{iq} samples $x$ pass through $\hdelay$ (for an $n$-cycle delay), $\htxzero$ (transmit channel), $\hchzero$ (wireless channel), and $\hrx$ (receive channel).
With only the antenna path active, the received signal is $\rzero$, which represents the \gls{si} signal.
In the cancellation path, $x$ passes through $\hfilt$ (\gls{ansic} filter), $\htxone$, and $\hchone$, yielding $\rone$ without $\hfilt$ and $\rfilt$ with it.
When both paths are active with the \gls{ansic} filter, the received signal is given as $\rcanc$
\begin{equation}\label{eq:canc_channel}
  \rcanc =  \hrx \ast \left(x \ast \hdelay \ast h_0 + x \ast \hfilt \ast h_1 \right) \, .
\end{equation}
where $\ast$ denotes convolution and $h_i = \htxi \ast \hchi$.
To fit $\hfilt$, we derive the \gls{lms} update rule for the $N$-tap filter $\boldhfilt$.
In vector-notation, for the $n$-th discrete time-step, the filter state is $\boldhfilt[n]$ and the measured residual \gls{si} is $\rcanc[n]$.
The task of the \gls{ansic} filter is to filter the transmitted baseband samples $\mathbf{x}[n] \in \C^{N}$ such that the transmitted signals from each port cancel each other out at an \gls{rf} combiner.
As a cost function, we use $C[n] = |\errorterm[n]|^2$ with $\errorterm[n] = \rcanc[n]$.
The gradient wrt. $\boldhfilt$ is given as $\nabla_{\boldhfilt} C[n] = \boldrone^*[n] \errorterm[n]$.
The filter update rule is
\begin{equation}\label{eq:canc_fitting_lms}
  \boldhfilt[n+1] = \boldhfilt[n] - \mu \nabla_{\boldhfilt} C[n] = \boldhfilt[n] - \mu \boldrone^*[n] \errorterm[n] \, ,
\end{equation}
where $\mu$ is the learning rate.
This update process utilizes $\rone$, rather than $x$, to account for the effect of the cancellation path.
Therefore, $\rone$ has to be measured.
This is done by transmitting only on the \gls{rf} port used for the cancellation and collecting the \gls{rx} baseband samples on the shared \gls{rx} port.
However, we later demonstrate that $\rone$ does not have to be measured often as the cancellation path is highly stable.

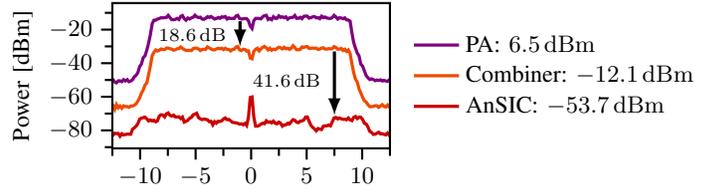
\begin{figure}[t]
  \centering
  \figuretextsize
  \centering
  \begin{tikzpicture}
    \begin{axis}[
      pgfstyledefault,
      width=\figurewidthsixty,
      height=\figureheightsixty,
      xmin=-12.5, xmax=12.5,
      xtick distance=5,
      ylabel={Power [dBm]},
      minor y tick num=1,
      legend style={
        at={(1.05, 0.5)},  % Adjust the x and y to control the position
        anchor=west,
        draw=none,
        fill=none,
        font=\figuretextsize,
      },
      legend cell align=left,  % Ensures left-aligned text in the legend
    ]

    \addplot[color=\Colorcoffpoffdoff, very thick] table [x=freq, y=pa_out, col sep=comma] {csv/canc_spectrum_analyzer.csv};
    \addlegendentry{PA: \SI{6.5}{\dBm}};

    \addplot[color=\Colorcoffpondon, very thick] table [x=freq, y=splitter_out, col sep=comma] {csv/canc_spectrum_analyzer.csv};
    \addlegendentry{Combiner: \SI{-12.1}{\dBm}};

    \addplot[color=\Colorconpoffdoffgoff, very thick] table [x=freq, y=canc_out, col sep=comma] {csv/canc_spectrum_analyzer.csv};
    \addlegendentry{\gls{ansic}: \SI{-53.7}{\dBm}};

    \draw[-latex, very thick] (axis cs:-1, -15) -- (axis cs:-1,-30);
    \node[left] at (axis cs:-1.5,-23) {\scriptsize \SI{18.6}{\decibel}};

    \draw[-latex, very thick] (axis cs:7.5,-33) -- (axis cs:7.5,-71);
    \node[left] at (axis cs:7,-50) {\scriptsize \SI{41.6}{\decibel}};

    \end{axis}
  \end{tikzpicture}\vspace*{-\subfigurecaptionspacesmall}
  \caption{Power measurements at different stages of the RF chain. The carrier frequency is \SI{2.472}{\giga\hertz}.}
  \label{fig:frequency_view}
\end{figure}

%%%%%%%%%%%%%%%%%%%%%%%%%%%%%%%%%%%%%%%%%%%%%%%%%%%%%%%%%%%%%%%%%%%%%%%%%%%%%%%
\section{Analog Self-Interference Cancellation}\label{sec:cancellation}

In this section, we first describe the hardware used for our \gls{ansic} filter.
Then, we present its performance.

%==============================================================================
\subsection{Hardware for Self-Interference Cancellation and Sensing}\label{sec:hardware_setup_for_sic_sensing}

Our testbed is based on a modified version of the openwifi project, an open-source IEEE 802.11 Wi-Fi \gls{sdr} implemented on a Xilinx XC7Z020-CLG484 \gls{fpga}~\cite{openwifigithub, jiaoOpenwifiFreeOpensource2020, kristensenMonostaticMultiTargetWiFiBased2024}.
The AD-FMCOMMS2-EBZ \gls{rf} front-end provides two \gls{rx} and two \gls{tx} chains.
For monostatic sensing, the system transmits and receives simultaneously, with a shared local oscillator for the \gls{tx} and \gls{rx} paths, eliminating the \gls{cfo}, \gls{sto}, and \gls{sfo} impairments.
The shared \gls{tx}/\gls{rx} antenna path includes a CN0417 \gls{pa} board ($\sim$\SI{20}{\decibel} gain)~\cite{analog_cn0417}, a MECA CS-2.5000 circulator ($\sim$\SI{13}{\decibel} attenuation)~\cite{meca_circulator}, a Mini-Circuits VBF-2435+ bandpass filter (\SI{2340}-\SI{2530}{\mega\hertz})~\cite{minicircuits_bpf}, and a Mini-Circuits ZX10-2-252-S+ splitter~\cite{minicircuits_splitter} to combine the paths for the shared \gls{adc}.
%The \gls{tx} power is $\sim$\SI{6.5}{\dBm} at the \gls{pa} output.
The \gls{ofdm} signal bandwidth is \SI{20}{\mega\hertz}.
This setup requires only one additional \gls{rf} port, the auxiliary \gls{tx} port, and uses off-the-shelf components, eliminating the need for custom \gls{pcb} and electronics design.
In contrast, similar monostatic systems with \gls{ansic}~\cite{barnetoFullDuplexOFDMRadar2019, hassaniInBandFullDuplexRadarCommunication2021, hassaniJointInBandFullDuplex2022, chenISACoTIntegratingSensing2022, chenISACFiEnablingFullfledged2024} often require sophisticated custom analog hardware and \glspl{pcb}.

%==============================================================================
\subsection{\gls{ansic} Filter Area Consumption}\label{sec:cancellation_area_consumption}

Table~\ref{tab:fpga_area} presents the \gls{fpga} area consumption separately for the \gls{ansic} filter and the openwifi system.
The \gls{ansic} filter processes 16-bit \gls{iq} samples using 16-bit weights, with a filter length of 20.
The system operates with a \SI{40}{\mega\hertz} clock.
While the \gls{fpga} is small, both the openwifi system and the \gls{ansic} filter fit, enabling the use of small and relatively cheap \glspl{fpga}.
Note that the \gls{ansic} filter is fitted on the CPU.

%==============================================================================
\subsection{Signal Spectrum Characteristics with \gls{ansic}}\label{sec:cancellation_performance_frequency_domain}

We validate the \gls{ansic} performance using measurements obtained using a Rohde \& Schwarz FS315~\cite{FS315Manual} spectrum analyzer.
For the measurements, we use a \gls{tx} port memory buffer to continuously transmit a Wi-Fi frame with no gaps in between the frame transmissions to ensure accurate power measurements.
The buffer contains a Wi-Fi frame of 3920 \gls{iq} samples, which is also used for \gls{ls} fitting.
\gls{rx} and cancellation port gains are pre-calibrated and the \gls{agc} disabled.

Fig.~\ref{fig:frequency_view} shows the signal spectrum at the \gls{pa} output after the \gls{rf} combiner with and without the \gls{ansic} filter.
Without \gls{ansic}, the power level at the \gls{rf} combiner output is \SI{-12.1}{\dBm}.
After applying \gls{ansic} with a \gls{ls}-fitted filter, the power reduces to \SI{-53.7}{\dBm}, indicating a \SI{41.6}{\decibel} cancellation.
The residual signal after \gls{ansic} shows $\sim$\SI{10}{\decibel} variation across the \SI{20}{\mega\hertz} bandwidth, indicating low frequency selectivity.
This contrasts with the \gls{ebd} setup in~\cite{hassaniInBandFullDuplexRadarCommunication2021}, which exhibits up to \SI{20}{\decibel} variation across the same bandwidth.
As noted in~\cite{laughlinWidelyTunableFull2015}, \glspl{ebd} generally show higher frequency selectivity than other cancellation techniques.
The systems in~\cite{chenISACFiEnablingFullfledged2024} and~\cite{barnetoFullDuplexOFDMRadar2019} demonstrate similarly low frequency selectivity to our results.
Our setup can achieve $\sim$\SI{42}{\decibel}, aligning well with the literature~\cite{barnetoFullDuplexOFDMRadar2019, hassaniInBandFullDuplexRadarCommunication2021, hassaniJointInBandFullDuplex2022, chenISACoTIntegratingSensing2022, chenISACFiEnablingFullfledged2024}.
With the \SI{13}{\decibel} of isolation from the circulator, and not accounting for the bandpass filter and \gls{rf} combiner, we achieve \SI{55}{\decibel} of isolation.

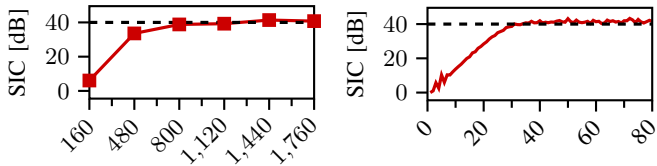
\begin{figure}[t]
  \centering
  \figuretextsize
\begin{subfigure}[t]{0.48\columnwidth}
  \centering
  \begin{tikzpicture}
    \begin{axis}[
      pgfstyledefault,
      width=\figurewidthfiftytwo,
      height=\figureheightfifty,
      xtick={160, 480, 800, 1120, 1440, 1760},
      xticklabel style={rotate=45, anchor=east, yshift=-0.4em},
      ylabel={SIC [dB]},
      ymin=0, ymax=45,
    ]

    \addplot[color=\Colorconpoffdoffgoff, mark=square*, very thick]
    coordinates {
      (160, 6.068542417336936)
      (480, 33.57230421840271)
      (800, 38.81702665902858)
      (1120, 39.25201208059721)
      (1440, 41.379411868559174)
      (1760, 40.759304942921304)
%      (2080, 41.64940391529932)
%      (2400, 42.62616242850697)
%      (2720, 41.05811757962168)
%      (3040, 42.79301011481676)
%      (3360, 42.91177285150367)
%      (3680, 40.971656662799475)
    };

%    \addplot[black, dashed, very thick] coordinates {(160, 40) (3680, 40)};
    \addplot[black, dashed, very thick] coordinates {(160, 40) (1760, 40)};
    \end{axis}
  \end{tikzpicture}\vspace*{-\subfigurecaptionspace}
  \caption{\gls{ls} \gls{sic} vs. \gls{iq} sample count.}
  \label{fig:cancellation_vs_data_size_ls}
\end{subfigure}\hfill
\begin{subfigure}[t]{0.48\columnwidth}
  \centering
  \figuretextsize
  \begin{tikzpicture}
    \begin{axis}[
      pgfstyledefault,
      width=\figurewidthfiftytwo,
      height=\figureheightfifty,
      xmin=0, xmax=80,
      xtick={0, 20, 40, 60, 80},
      xticklabels={0, 20, 40, 60, \phantom{1,7}80},
      xticklabel style={rotate=45, anchor=east, yshift=-0.4em},
      ylabel={SIC [dB]},
      ymin=0, ymax=45,
    ]

    \addplot[color=\Colorconpoffdoffgoff, very thick] table [x expr=\coordindex, y=cancellation, col sep=comma] {csv/train_time/one_pa/cancellation_data.csv};

    \addplot[black, dashed, very thick] coordinates {(0, 40) (200, 40)};

    \end{axis}
  \end{tikzpicture}\vspace*{-\subfigurecaptionspace}
  \caption{\gls{sic} vs. number of \gls{lms} steps.}
  \label{fig:cancellation_vs_train_steps_lms}
\end{subfigure}
  \vspace*{\subfigurecaptionspacesmall}
  \caption{\gls{sic} as a function of (a) the number of \gls{iq} samples for \gls{ls} and (b) the number of \gls{lms} steps, respectively.}
\end{figure}

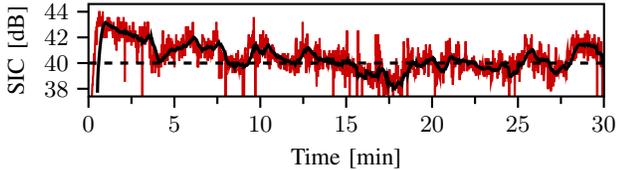
\begin{figure}[t]
  \centering
  \figuretextsize
  \begin{tikzpicture}
    \begin{axis}[
      pgfstyledefault,
      width=\figurewidthninetyfive,
      height=\figureheightfifty,
      xlabel={Time [min]},
      xtick distance=5,
      ylabel={SIC [dB]},
      ymin=38, ymax=44,
      ytick distance=2,
      legend style={at={(0.5,1.35)}, anchor=north, legend columns=-1},
    ]

    \addplot[color=\Colorconpoffdoffgoff, thick] table [x=timestamp, y=cancellation, col sep=comma] {csv/stability/overall_cancellation_data.csv};
    \addplot[color=black, very thick] table [x=timestamp, y=running_avg_cancellation, col sep=comma] {csv/stability/overall_cancellation_data.csv};
    \addplot[black, dashed, very thick] coordinates {(0, 40) (30, 40)};

    \end{axis}
  \end{tikzpicture}
  \caption{\textcolor{\Colorconpoffdoffgoff}{Red} indicates the \SI{10}{\hertz} \gls{sic} measurements and \textcolor{black}{black} the \gls{sic} moving average over \SI{30}{\second}.}
  \label{fig:stability_error_plots}
\end{figure}

%==============================================================================
\subsection{\gls{ansic} Performance for LS and LMS Fitting}\label{sec:cancellation_performance_data_size}

This section evaluates the requirements in terms of number of \gls{iq} samples for \gls{ls} and \gls{lms} to fit \gls{ansic} filter weights.
Fig.~\ref{fig:cancellation_vs_data_size_ls} illustrates the \gls{sic} improvement as a function of the number of \gls{iq} samples used for \gls{ls} fitting.
The \gls{sic} significantly improves from 160 to 480 samples and plateaus around 1440 samples (\SI{72}{\micro\second} of air-time).
\SI{40}{\decibel} of \gls{sic} is achieved with 800 samples (\SI{40}{\micro\second} of air-time), indicating high \gls{sic} can be achieved with short Wi-Fi frame segments.
While \gls{ls} is effective for initial training, \gls{lms} is more suitable for channel tracking.
For \gls{lms} fitting, 80 \gls{iq} samples from the \gls{htltf} field are used, representing \SI{4}{\micro\second}.
The same $\rone$ measurement is reused across all training iterations.
Fig.~\ref{fig:cancellation_vs_train_steps_lms} shows convergence in about 40 \gls{lms} steps (\SI{128}{\micro\second} of air-time).
When executed on the CPU, the training rate can reach up to $\sim$\SI{500}{\hertz} when fitting based on $80$ \gls{iq} samples.
The reuse of $\rone$ for several \gls{lms} steps underscores the cabled cancellation path stability and demonstrates that it is not necessary to re-measure $\rone$ for every filter update, simplifying the process of fine-tuning.

%==============================================================================
\subsection{Long-Term Stability of \gls{ansic}}\label{sec:cancellation_performance_stability}

\andreas{For very dense time-series plots, just use thick, add as note somewhere like in color sceme}
\andreas{Properly align these 2 figures. Not sure if it's the y expand option or what it is}
\andreas{In principle, if you had like 10 minutes of data, you could probably estimate the breathing without SI interference, but it would be doubltful if that is too long for human, because we have a minimum of data we need to estimate? Not sure, is 2 minutes needed to clean it up?}

A 30 minute experiment was conducted in a populated office environment to evaluate the \gls{ansic} stability.
Following initial \gls{lms} fitting, \gls{rx} power measurements were collected at \SI{10}{\hertz} on the openwifi board.
Fig.~\ref{fig:stability_error_plots} illustrates the ability of our system to maintain an average \gls{sic} of $\sim$\SI{40}{\decibel} throughout most of the 30 minute experiment.
This consistent performance indicates stable \gls{tx} paths after initial fitting, with no significant drifts between the two \gls{tx} paths.
A brief \gls{sic} degradation occurred around \SI{15}{\minute}, but the system recovered to \SI{42}{\decibel} cancellation by the end of the experiment.
Our system thus maintains robust cancellation despite occasional interference.
The work demonstrates potential for long-term \gls{sic} stability without frequent retraining, contrasting with monostatic setups in~\cite{hassaniInBandFullDuplexRadarCommunication2021, hassaniJointInBandFullDuplex2022} that continuously retrain, and~\cite{barnetoFullDuplexOFDMRadar2019, chenISACoTIntegratingSensing2022, chenISACFiEnablingFullfledged2024} where the retraining frequency is unclear or unspecified during sensing.

%%%%%%%%%%%%%%%%%%%%%%%%%%%%%%%%%%%%%%%%%%%%%%%%%%%%%%%%%%%%%%%%%%%%%%%%%%%%%%%
\section{Results - Sensing}\label{sec:results}

In this section, we evaluate our system for sensing.
We discuss two approaches: first we adapt the filter weights continuously and use the changing weights for sensing.
Then, we fix the \gls{ansic} weights and perform sensing based on the residual signal.

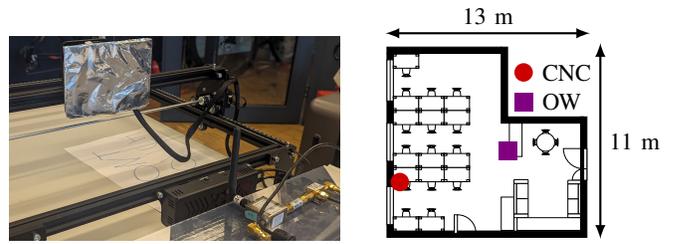
\begin{figure}[t]
  \centering
  \figuretextsize
  \begin{subfigure}[t]{0.45\columnwidth} % \andreas{Make all of these consistent}
    \centering
    \includegraphics[trim={20cm 15cm 10cm 25cm},clip, width=\figurewidthfifty]{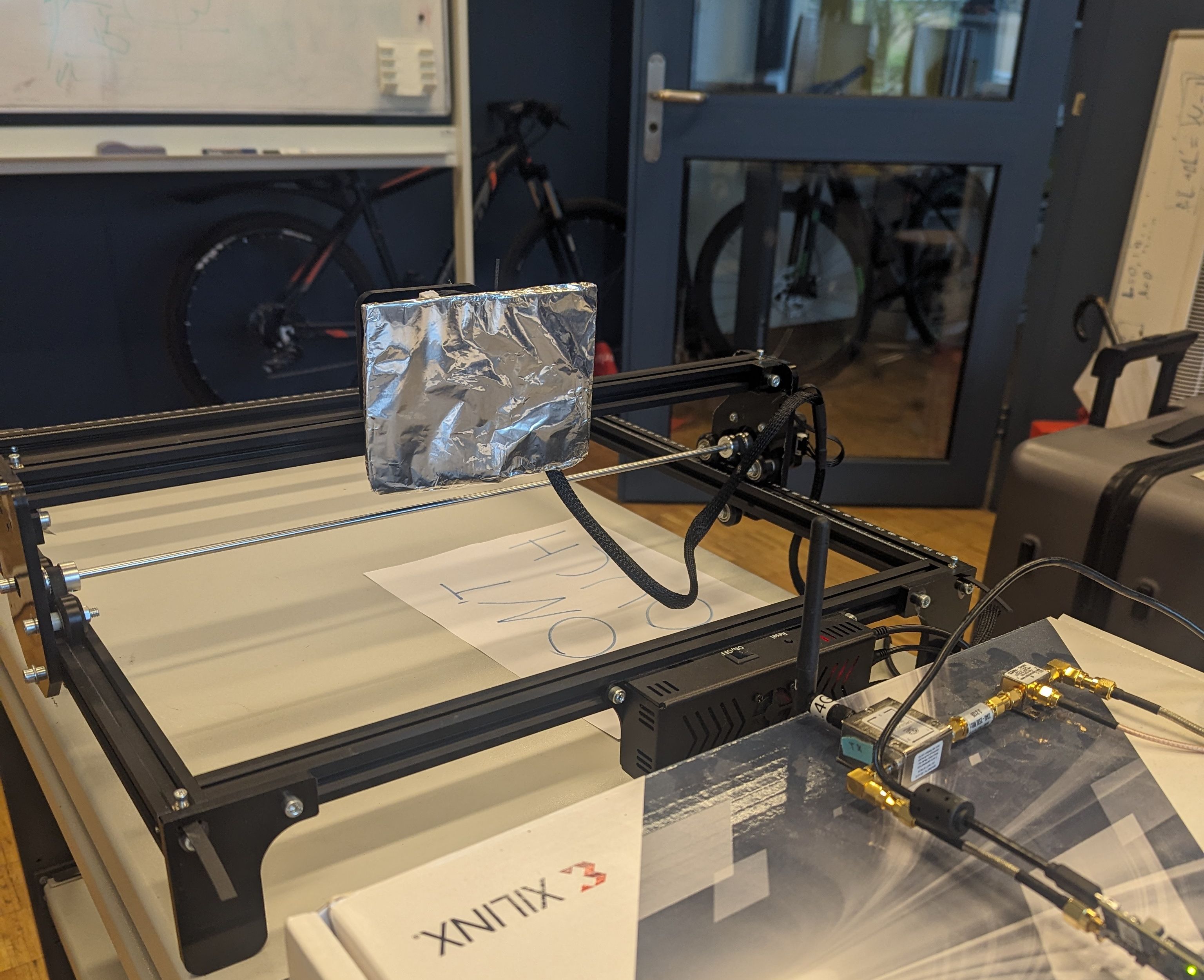}
    \caption{CNC-machine for \gls{ansic} filter weight sensing.}
    \label{fig:lms_experimental_setup}
  \end{subfigure}\hfill
  \begin{subfigure}[t]{0.45\columnwidth}
    \centering
\begin{tikzpicture}[
  locationA/.style={circle, minimum size=7pt, inner sep=0pt, outer sep=0pt, fill=\Colorconpoffdoffgoff, text=white, font=\bfseries},
  locationB/.style={rectangle, minimum size=7pt, inner sep=0pt, outer sep=0pt, fill=\Colorcoffpoffdoff, text=white, font=\bfseries},
]
  \node[anchor=south west,inner sep=0] (image) at (0,0) {\includegraphics[width=0.65\linewidth, angle=90]{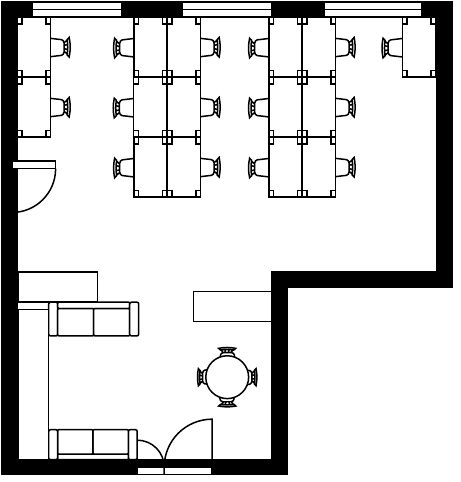}};
  \begin{scope}[x={(image.south east)},y={(image.north west)}]
    % Define coordinates
    \coordinate (CNC2) at (0.07,0.29);
    \coordinate (WiFi1) at (0.6,0.45);
    % Add CNC machine locations (red circles) with white numbers inside
    \node[locationA] at (CNC2) {};
    \node[locationB] at (WiFi1) {};
    % Room dimensions
    \draw[<->,thick] (1.05,0) -- node[right] {11 m} (1.05,1);
    \draw[<->,thick] (0,1.05) -- node[above] {13 m} (1,1.05);
    % Add legend
    \node[locationA] at (0.675, 0.85) {};
    \node[anchor=west] at (0.725, 0.85) {CNC};
    \node[locationB] at (0.675, 0.7) {};
    \node[anchor=west] at (0.725, 0.7) {OW};
  \end{scope}
\end{tikzpicture}
    \caption{Floorplan with openwifi (OW) and \gls{cnc} location.}
    \label{fig:room_floorplan}
  \end{subfigure}
  \caption{Photo and floorplan of the setup used for the \gls{ansic} filter sensing test (a) and the \gls{nlos} \gls{cir} sensing test (b).}
  \label{fig:room_setup}
\end{figure}

%==============================================================================
\subsection{Sensing with \gls{ansic} Filter Weights}\label{sec:cancellation_performance_sensing}

When $\boldhfilt$ is fine-tuned to the channel, it contains information on all components in~\eqref{eq:model_contributors}, including the dynamic ($\gd$) ones.
We demonstrate this using the experimental setup in Fig.~\ref{fig:lms_experimental_setup}, which uses a \gls{cnc} machine with an aluminum foil covered plate ($8\,\mathrm{cm} \times 12\,\mathrm{cm}$) to emulate breathing.
The \gls{cnc} machine moves \SI{1}{\centi\meter} at 12 \gls{bpm} to emulate human breathing~\cite{boussugesDiaphragmaticMotionStudied2009} and \SI{20}{\centi\meter} to evaluate performance under stronger interference, aligning with related works~\cite{barnetoFullDuplexOFDMRadar2019, hassaniInBandFullDuplexRadarCommunication2021, hassaniJointInBandFullDuplex2022, chenISACoTIntegratingSensing2022, chenISACFiEnablingFullfledged2024} that focus on larger and faster-moving targets.
\SI{12}{\bpm} is chosen as it corresponds to the smallest value in the \SI{12} to \SI{15}{\bpm} range for human breathing at rest~\cite[p. 588]{barrettGanongReviewMedical2010}.

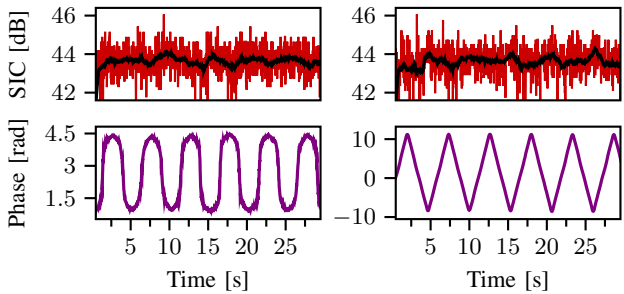
\begin{figure}[t]
  \centering
  \figuretextsize
%  \begin{subfigure}[t]{\columnwidth}
%    \centering
    \begin{tikzpicture}
      \begin{groupplot}[
        pgfstyledefault,
        group style={
          group size=2 by 2,
          horizontal sep=1cm,
          vertical sep=0.35cm,
        },
        width=\figurewidthfiftytwo,
        height=\figureheightfifty,
        xmin=0.6, xmax=29.4
      ]

      % Other
      % 10mm: u16, f0.9, h0.9 or u15 f0 h0
      % 200mm: u16 f0.5 h0.9 or u15 f0 h0
      \nextgroupplot[ylabel={SIC [dB]}, y label style={anchor=south, at={(-0.26, 0.5)}}, ymin=42, ymax=46, xtick=\empty]
      \addplot[color=\Colorconpoffdoffgoff, thick] table [x=timestamp, y=cancellation, col sep=comma] {csv/lms/results_10mm/m0_i0_e1_t30_v0_u15_n20_s6_d80_r100_o640_w1_g0_a0_f0p000000_c0_l0_h0p000000_20240502_221735_cancellation_data.csv};

      \addplot[color=black, very thick] table [x=timestamp, y=smoothed_cancellation, col sep=comma] {csv/lms/results_10mm/m0_i0_e1_t30_v0_u15_n20_s6_d80_r100_o640_w1_g0_a0_f0p000000_c0_l0_h0p000000_20240502_221735_cancellation_data.csv};

      \nextgroupplot[ymin=42, ymax=46, xtick=\empty]
      \addplot[color=\Colorconpoffdoffgoff, thick] table [x=timestamp, y=cancellation, col sep=comma] {csv/lms/results_200mm/m0_i0_e1_t30_v0_u15_n20_s6_d80_r100_o640_w1_g0_a0_f0p000000_c0_l0_h0p000000_20240502_233436_cancellation_data.csv};

      \addplot[color=black, very thick] table [x=timestamp, y=smoothed_cancellation, col sep=comma] {csv/lms/results_200mm/m0_i0_e1_t30_v0_u15_n20_s6_d80_r100_o640_w1_g0_a0_f0p000000_c0_l0_h0p000000_20240502_233436_cancellation_data.csv};

      % Feature plots
% , ymin=1, ymax=5
      \nextgroupplot[xlabel={Time [s]}, xmin=0.5, xmax=29.5, xtick distance=5, ylabel={Phase [rad]}, ytick distance=1.5]
%      \nextgroupplot[xlabel={Time [s]}, xmin=0.5, xmax=29.5, xtick distance=5, ylabel={Phase [rad]}]
      \addplot[color=\Colorcoffpoffdoff, very thick] table [x=timestamp, y=w6, col sep=comma] {csv/lms/results_10mm/m0_i0_e1_t30_v0_u15_n20_s6_d80_r100_o640_w1_g0_a0_f0p000000_c0_l0_h0p000000_20240502_221735_phase_zero_mean_data.csv};

      \nextgroupplot[xlabel={Time [s]}, xmin=0.5, xmax=29.5, xtick distance=5]
      \addplot[color=\Colorcoffpoffdoff, very thick] table [x=timestamp, y=w6, col sep=comma] {csv/lms/results_200mm/m0_i0_e1_t30_v0_u15_n20_s6_d80_r100_o640_w1_g0_a0_f0p000000_c0_l0_h0p000000_20240502_233436_phase_zero_mean_data.csv};

%      \nextgroupplot[ylabel={SIC [dB]}, y label style={anchor=south, at={(-0.26, 0.5)}}, ymin=42, ymax=46, xtick=\empty]
%      \addplot[color=\Colorconpoffdoffgoff, thick] table [x=timestamp, y=cancellation, col sep=comma] {csv/lms/results_cnc_10mm_12bpm_cancellation_data.csv};
%
%      \nextgroupplot[ymin=42, ymax=46, xtick=\empty]
%      \addplot[color=\Colorconpoffdoffgoff, thick] table [x=timestamp, y=cancellation, col sep=comma] {csv/lms/results_cnc_200mm_12bpm_cancellation_data.csv};
%
%      % Feature plots
%      \nextgroupplot[xlabel={Time [s]}, xmin=0.5, xmax=29.5, xtick distance=5, ylabel={Phase [rad]}, ytick distance=1.5, ymin=1, ymax=5]
%      \addplot[color=\Colorcoffpoffdoff, very thick] table [x=timestamp, y=w6, col sep=comma] {csv/lms/results_cnc_10mm_12bpm_phase_data.csv};
%
%      \nextgroupplot[xlabel={Time [s]}, xmin=0.5, xmax=29.5, xtick distance=5]
%      \addplot[color=\Colorcoffpoffdoff, very thick] table [x=timestamp, y=w6, col sep=comma] {csv/lms/results_cnc_200mm_12bpm_phase_data.csv};

      \end{groupplot}
    \end{tikzpicture}
  \caption{\gls{sic} (top) and filter weight (zero-meaned) phase (bottom) with \gls{lms} and CNC movement. \SI{1}{\second} \gls{sic} moving average.}
  \label{fig:ansic_results}
\end{figure}

The \gls{sic} results in Fig.~\ref{fig:ansic_results} show that our \gls{ansic} filter maintains $\sim$\SI{44}{\decibel} \gls{sic} in both scenarios, indicating effective real-time adaptation.
While real-time adaption has also been demonstrated in~\cite{hassaniInBandFullDuplexRadarCommunication2021, hassaniJointInBandFullDuplex2022} where the \gls{sic} is stable under close movements, we further demonstrate the ability to perform sensing based on the filter coefficients as shown in Fig.~\ref{fig:ansic_results}.
The phase of the main \gls{ansic} filter weight clearly represents the periodic movement of the \gls{cnc} machine, indicating real-time adaptation to changing interference patterns.
While~\cite{hassaniJointInBandFullDuplex2022} also demonstrates a similar use of cancellation filter weights for sensing, they use a digital canceller.

\begin{figure}[t]
  \centering
  \figuretextsize
\begin{subfigure}[t]{\columnwidth}
  \begin{tikzpicture}
    \begin{axis}[
      pgfstyleonedspectrum,
      ymin=-75, ymax=-45, % 30 dB range
      ytick={-70, -60, -50},
      legend style={at={(0.5, 1.0)}, anchor=south, legend columns=-1, draw=none, fill=none, font=\figuretextsize},
    ]

    \addplot[color=\Colorcoffpoffdoff, very thick] table [x=freq, y=feature5, col sep=comma] {csv/sensing/cir_spectrum_2d_main_bins_canc_off_lab_nlos_10mm_12bpm_full_spectrum2d_plot_non_normalized.csv};
    \addlegendentry{No \gls{ansic}};

    \addplot[color=\Colorconpoffdoffgon, very thick] table [x=freq, , y expr=\thisrow{feature5} - 42, col sep=comma] {csv/sensing/cir_spectrum_2d_main_bins_canc_on_gain_incr_lab_nlos_10mm_12bpm_full_spectrum2d_plot_non_normalized.csv};
    \addlegendentry{\gls{ansic} + gain};

    \addplot[black, densely dashed, thick] coordinates {(11.91, -100) (11.91, 15)};

    \end{axis}
  \end{tikzpicture}
\end{subfigure}\vspace*{-\subfigurevstackvspace}
  \caption{Frequency spectrum for the \SI{50}{\nano\second} \gls{cir} delay bin.}
  \label{fig:nlos1_1cm_freq_1d}
  \vspace*{\subfigurevstackvspace}
  \vspace*{\subfigurevstackvspace}
  \vspace*{\subfigurevstackvspace}
  \vspace*{\subfigurevstackvspace}

\begin{subfigure}[t]{\columnwidth}
  \centering
  \figuretextsize
  \begin{tikzpicture}
    \begin{groupplot}[
      pgfstyledefault,
      group style={
        group size=2 by 1,
        horizontal sep=\groupplotsep,
        vertical sep=\groupplotsep,
      },
      width=\figurewidthsixtyfive,
      height=\figureheightfifty,
      xlabel={Time [s]},
      xmin=0.5, xmax=29.5,
      xtick distance=5,
      ylabel={},
      ytick=\empty,
      legend style={at={(0.5, 1.0)}, anchor=south, draw=none, fill=none},
    ]

    \nextgroupplot[ylabel={$|g|$}, y label style={anchor=south, at={(-.05, 0.5)}}]
    \addplot[color=\Colorcoffpoffdoff, very thick, restrict x to domain=0:30] table [x=Timestamp, y=FilteredMagnitude, col sep=comma] {csv/sensing/cir_spectrum_2d_main_bins_canc_off_lab_nlos_10mm_12bpm_signal5.csv};
    \addlegendentry{No \gls{ansic}};

    \nextgroupplot
    \addplot[color=\Colorconpoffdoffgon, very thick, restrict x to domain=0:30] table [x=Timestamp, y=FilteredMagnitude, col sep=comma] {csv/sensing/cir_spectrum_2d_main_bins_canc_on_gain_incr_lab_nlos_10mm_12bpm_signal5.csv};
    \addlegendentry{\gls{ansic} + gain};

    \end{groupplot}
  \end{tikzpicture}
\end{subfigure}\vspace*{-\subfigurevstackvspace}
  \caption{Magnitude for the \SI{50}{\nano\second} \gls{cir} delay bin over \SI{30}{\second}.}
  \label{fig:nlos1_1cm_time}
\end{figure}

%==============================================================================
\subsection{CIR-Based Sensing}\label{sec:results_sensing_cir}

To perform sensing on the residual \gls{si}, we also consider the \gls{cir} estimated from the residual received signal when the \gls{ansic} filter is fixed after initial \gls{ls}-fitting.
\gls{cir}-based sensing experiments were conducted in a lab space (Fig.~\ref{fig:room_floorplan}) using the same \gls{cnc} machine as before (Fig.~\ref{fig:lms_experimental_setup}) to again emulate human breathing.
To make detection more challenging, we consider a \gls{nlos} scenario, with the openwifi board and \gls{cnc} locations in Fig.~\ref{fig:room_floorplan}.
This scenario has a \SI{10}{\meter} direct path, but the target is not visible from the position of the openwifi board and is hidden behind a lab desk setup.
The \gls{cnc} machine moves \SI{1}{\centi\meter} at \SI{12}{\bpm} to emulate breathing.

For \gls{ansic}, the board was configured as described in Section~\ref{sec:hardware_setup_for_sic_sensing}.
When \gls{ansic} is disabled, the cancellation port is attenuated to prevent noise through the combiner.
Wi-Fi frames of 3920 \gls{iq} samples were transmitted at \SI{100}{\hertz} for 1~minute.
The \gls{rx} \gls{iq} collection is triggered by the transmission of the first sample from the \gls{fpga}.
To estimate the \gls{cir}, we extracted all \gls{ofdm} symbols from the \gls{iq} samples, computed the \gls{cir} for each \gls{ofdm} symbol, and then averaged these across all symbols within each frame.
The data was resampled to \SI{10}{\hertz} for a uniform sampling rate.
As the \gls{ofdm} bandwidth is \SI{20}{\mega\hertz}, the \gls{cir} delay resolution is \SI{25}{\nano\second} or \SI{7.5}{\meter}.
We define the \SI{0}{\nano\second} delay bin as the one with the highest power, representing the direct path \gls{si}.
As the target is at distance of \SI{10}{\meter} in \gls{nlos} conditions, we expect the target to show up around the \SI{25}{\nano\second} or \SI{50}{\nano\second} delay bins, corresponding to distances of \SI{7.5}{\meter} and \SI{15}{\meter}.
When analyzing all delay bins in the frequency domain, the \SI{50}{\nano\second} bin was found to be the strongest at \SI{12}{\bpm}.
Hence, we focus on this bin.

\subsubsection*{Frequency-Domain}
Fig.~\ref{fig:nlos1_1cm_freq_1d} shows the spectrum of the \SI{50}{\nano\second} bin for \SI{1}{\minute} of measurements.
For the frequency spectra, the curve is shifted down for the \gls{rx} gain increase.
Without \gls{ansic}, no distinct peak at \SI{12}{\bpm} is observed, indicating difficulty in estimating slow targets under \gls{si}.
Applying \gls{ansic} recovers a strong peak at \SI{12}{\bpm} and enables an \gls{rx} gain increase of $\sim$\SI{40}{\decibel} from the \gls{rx} power reduction.
These findings highlight the challenges of \gls{nlos} sensing and the necessity of \gls{si} mitigation techniques, with \gls{ansic} to effectively reduce the impact of \gls{si} and optimize \gls{rx} gain and \gls{adc} utilization.

\subsubsection*{Time-Domain}

Fig.~\ref{fig:nlos1_1cm_time} shows the magnitude of the \SI{50}{\nano\second} bin.
No clear periodic signal is visible without \gls{ansic}.
With \gls{ansic}, Fig.~\ref{fig:nlos1_1cm_time} reveals a clear signal, although some distortion remains.
These results, in alignment with the frequency domain findings, demonstrate that with \gls{ansic}, sensing targets can be recovered at a distance of \SI{10}{\meter} in \gls{nlos} conditions.

%%%%%%%%%%%%%%%%%%%%%%%%%%%%%%%%%%%%%%%%%%%%%%%%%%%%%%%%%%%%%%%%%%%%%%%%%%%%%%%
\section{Conclusion}

In this paper, we presented an \gls{sdr}-based monostatic Wi-Fi sensing system that employs an auxiliary \gls{ansic} approach.
Our system achieves \gls{sic} levels comparable to existing solutions while simplifying hardware requirements.
We have demonstrated that unlike previous works~\cite{barnetoFullDuplexOFDMRadar2019, hassaniInBandFullDuplexRadarCommunication2021, hassaniJointInBandFullDuplex2022, chenISACoTIntegratingSensing2022, chenISACFiEnablingFullfledged2024} our setup preserves sensing information in two ways.
First, the \gls{ansic} filter weights can be used directly for sensing with \gls{lms} fine-tuning.
Second, the stability of the cancellation, allows for long periods without fine-tuning the filter while maintaining a high level of \gls{sic}.
This enables sensing from received \gls{iq} samples and allow for traditional vital sign sensing using channel estimates derived from baseband samples without interference from the canceller's adaptation -- a significant improvement over previous systems that require careful consideration of the effect of the fine-tuning when performing sensing.
Our experiments confirmed the ability of our system to detect small, slow-moving targets at distances of up to \SI{10}{\meter} in \gls{nlos} conditions.
These findings indicate that effective \gls{sic} and accurate sensing in monostatic systems can be achieved without custom hardware and with stable performance.

%%%%%%%%%%%%%%%%%%%%%%%%%%%%%%%%%%%%%%%%%%%%%%%%%%%%%%%%%%%%%%%%%%%%%%%%%%%%%%%
\section*{Acknowledgment}
This research has been kindly supported by the Swiss National Science Foundation under Grant-ID 182621.

%%%%%%%%%%%%%%%%%%%%%%%%%%%%%%%%%%%%%%%%%%%%%%%%%%%%%%%%%%%%%%%%%%%%%%%%%%%%%%%
%\balance
%\bibliographystyle{IEEEtran}
%\bibliography{IEEEabrv, bibliography}

\balance
%\bstctlcite{IEEEexample:BSTcontrol}
\bibliographystyle{IEEEtran}

{\bibliography{bibliography, bib_general,bib_wifi_sensing,bib_fmcw_sensing,bib_uwb_sensing,bib_mmwave_sensing,bib_surveys,bib_jsac,bib_openwifi}}

\end{document}

%% file: macros.tex
\newcommand{\bmg}{\mathbf{g}}

\newcommand{\C}{\mathbb{C}}

%% file: colorscheme.tex
\definecolor{bostonunired}{HTML}{CC0000}  % https://www.color-name.com/hex/CC0000
\definecolor{fireenginered}{HTML}{CF202A} % https://www.color-name.com/hex/CF202A
\definecolor{vermillionred}{HTML}{E34234} % https://www.color-name.com/vermillion.color

\definecolor{titlered}{RGB}{212,0,0}

\definecolor{reddeep}{RGB}{178,24,43}
\definecolor{redlight}{RGB}{252,78,42}
\definecolor{redlight2}{RGB}{255,114,111}

\definecolor{redbright}{RGB}{255,24,43}

\definecolor{perfectgreen}{HTML}{4FBF26}  % https://www.color-name.com/perfect-green.color
\definecolor{maygreen}{HTML}{4E9B47}      % https://www.color-name.com/hex/4E9B47
\definecolor{mattelime}{HTML}{75AD50}     % https://www.color-name.com/matte-lime.color

\definecolor{indigorainbow}{HTML}{1D3F6E} % https://www.color-name.com/hex/1D3F6E
\definecolor{royalazure}{HTML}{1866E1}    % https://www.color-name.com/royal-azure.color
\definecolor{azure}{HTML}{008AFF}         % https://www.color-name.com/azure-traditional.color
\definecolor{blueberry}{HTML}{4F86F7}     % https://www.color-name.com/blueberry.color
\definecolor{fadednavy}{HTML}{242F78}     % https://www.color-name.com/faded-navy.color
\definecolor{navy}{HTML}{000080}          % https://www.color-name.com/navy.color

\definecolor{goldwebgolden}{HTML}{FFD700} % https://www.color-name.com/hex/FFD700
\definecolor{radioactive}{HTML}{FAE500}   % https://www.color-name.com/radioactive.color
\definecolor{gold}{HTML}{FFD700}          % https://www.color-name.com/gold.color

\definecolor{vividgamboge}{HTML}{FF9900} % https://www.color-name.com/hex/FF9900

\definecolor{shinygray}{HTML}{C7C6C6}

\definecolor{slategrey}{HTML}{708090}     % https://www.color-name.com/slate-gray.color
\definecolor{neutralgray}{HTML}{828382}   % https://www.color-name.com/neutral-grey.color
\definecolor{mattecharcoal}{HTML}{3B4248} % https://www.color-name.com/matte-charcoal.color
\definecolor{graydark}{HTML}{6B6E70}      % https://www.color-name.com/antique-steel.color

\definecolor{charcoal}{HTML}{36454F}        % https://www.color-name.com/charcoal.color
\definecolor{charcoalShade1}{HTML}{2B373F}
\definecolor{charcoalShade2}{HTML}{263037}
\definecolor{charcoalShade3}{HTML}{20292F}

\definecolor{sunset1}{HTML}{F3E79B}
\definecolor{sunset2}{HTML}{FAC484}
\definecolor{sunset3}{HTML}{F8A07E}
\definecolor{sunset4}{HTML}{EB7F86}
\definecolor{sunset5}{HTML}{CE6693}
\definecolor{sunset6}{HTML}{A059A0}
\definecolor{sunset7}{HTML}{5C53A5}

\definecolor{bluyi1}{HTML}{F7FEAE}
\definecolor{bluyi2}{HTML}{B7E6A5}
\definecolor{bluyi3}{HTML}{7CCBA2}
\definecolor{bluyi4}{HTML}{46AEA0}
\definecolor{bluyi5}{HTML}{089099}
\definecolor{bluyi6}{HTML}{00718B}
\definecolor{bluyi7}{HTML}{045275}

\definecolor{geyser1}{HTML}{008080}
\definecolor{geyser2}{HTML}{70A494}
\definecolor{geyser3}{HTML}{B4C8A8}
\definecolor{geyser4}{HTML}{F6EDBD}
\definecolor{geyser5}{HTML}{EDBB8A}
\definecolor{geyser6}{HTML}{DE8A5A}
\definecolor{geyser7}{HTML}{CA562C}

\definecolor{temps1}{HTML}{009392}
\definecolor{temps2}{HTML}{39B185}
\definecolor{temps3}{HTML}{9CCb86}
\definecolor{temps4}{HTML}{E9E29C}
\definecolor{temps5}{HTML}{EEB479}
\definecolor{temps6}{HTML}{E88471}
\definecolor{temps7}{HTML}{CF597E}

\definecolor{earth1}{HTML}{A16928}
\definecolor{earth2}{HTML}{BD925A}
\definecolor{earth3}{HTML}{D6BD8D}
\definecolor{earth4}{HTML}{EDEAC2}
\definecolor{earth5}{HTML}{B5C8B8}
\definecolor{earth6}{HTML}{79A7AC}
\definecolor{earth7}{HTML}{2887A1}

\definecolor{bold1}{HTML}{7F3C8D}
\definecolor{bold2}{HTML}{11A579}
\definecolor{bold3}{HTML}{3969AC}
\definecolor{bold4}{HTML}{F2B701}
\definecolor{bold5}{HTML}{E73F74}
\definecolor{bold6}{HTML}{80BA5A}
\definecolor{bold7}{HTML}{E68310}
\definecolor{bold8}{HTML}{008695}
\definecolor{bold9}{HTML}{CF1C90}
\definecolor{bold10}{HTML}{F97B72}
\definecolor{bold11}{HTML}{4B4B8F}
\definecolor{bold12}{HTML}{A5AA99}

\definecolor{pastel1}{HTML}{66C5CC}
\definecolor{pastel2}{HTML}{F6CF71}
\definecolor{pastel3}{HTML}{F89C74}
\definecolor{pastel4}{HTML}{DCB0F2}
\definecolor{pastel5}{HTML}{87C55F}
\definecolor{pastel6}{HTML}{9EB9F3}
\definecolor{pastel7}{HTML}{FE88B1}
\definecolor{pastel8}{HTML}{C9DB74}
\definecolor{pastel9}{HTML}{8BE0A4}
\definecolor{pastel10}{HTML}{B497E7}
\definecolor{pastel11}{HTML}{D3B484}
\definecolor{pastel12}{HTML}{B3B3B3}

\definecolor{prism1}{HTML}{5F4690}
\definecolor{prism2}{HTML}{1D6996}
\definecolor{prism3}{HTML}{38A6A5}
\definecolor{prism4}{HTML}{0F8554}
\definecolor{prism5}{HTML}{73AF48}
\definecolor{prism6}{HTML}{EDAD08}
\definecolor{prism7}{HTML}{E17C05}
\definecolor{prism8}{HTML}{CC503E}
\definecolor{prism9}{HTML}{94346E}
\definecolor{prism10}{HTML}{6F4070}
\definecolor{prism11}{HTML}{994E95}
\definecolor{prism12}{HTML}{666666}

\definecolor{vivid1}{HTML}{E58606}
\definecolor{vivid2}{HTML}{5D69B1}
\definecolor{vivid3}{HTML}{52BCA3}
\definecolor{vivid4}{HTML}{99C945}
\definecolor{vivid5}{HTML}{CC61B0}
\definecolor{vivid6}{HTML}{24796C}
\definecolor{vivid7}{HTML}{DAA51B}
\definecolor{vivid8}{HTML}{2F8AC4}
\definecolor{vivid9}{HTML}{764E9F}
\definecolor{vivid10}{HTML}{ED645A}
\definecolor{vivid11}{HTML}{CC3A8E}
\definecolor{vivid12}{HTML}{A5AA99}

\definecolor{cBrewerPaired1}{HTML}{A6CEE3}
\definecolor{cBrewerPaired2}{HTML}{1F78B4}
\definecolor{cBrewerPaired3}{HTML}{B2DF8A}
\definecolor{cBrewerPaired4}{HTML}{33A02C}
\definecolor{cBrewerPaired5}{HTML}{FB9A99}
\definecolor{cBrewerPaired6}{HTML}{E31A1C}
\definecolor{cBrewerPaired7}{HTML}{FDBF6F}
\definecolor{cBrewerPaired8}{HTML}{FF7F00}
\definecolor{cBrewerPaired9}{HTML}{CAB2D6}
\definecolor{cBrewerPaired10}{HTML}{6A3D9A}
\definecolor{cBrewerQualPrint1}{HTML}{E41A1C}
\definecolor{cBrewerQualPrint2}{HTML}{377EB8}
\definecolor{cBrewerQualPrint3}{HTML}{4DAF4A}
\definecolor{cBrewerQualPrint4}{HTML}{984EA3}
\definecolor{cBrewerQualPrint5}{HTML}{FF7F00}
\definecolor{cBrewerQualPrint6}{HTML}{FFFF33}
\definecolor{cBrewerQualPrint7}{HTML}{A65628}
\definecolor{cBrewerQualPrint8}{HTML}{F781BF}
\definecolor{cBrewerQualPrint9}{HTML}{999999}

\pgfplotsset{colormap={inferno}{
  rgb(0)=(0.001462, 0.000466, 0.013866),
  rgb(15)=(0.037668, 0.025921, 0.132232),
  rgb(30)=(0.116656, 0.047574, 0.272321),
  rgb(45)=(0.217949, 0.036615, 0.383522),
  rgb(60)=(0.316282, 0.053490, 0.425116),
  rgb(75)=(0.410113, 0.087896, 0.433098),
  rgb(90)=(0.503493, 0.121575, 0.423356),
  rgb(105)=(0.596940, 0.154848, 0.398125),
  rgb(120)=(0.688653, 0.192239, 0.357603),
  rgb(135)=(0.775059, 0.239667, 0.303526),
  rgb(150)=(0.851384, 0.302260, 0.239636),
  rgb(165)=(0.912966, 0.381636, 0.169755),
  rgb(180)=(0.956852, 0.475356, 0.094695),
  rgb(195)=(0.981895, 0.579392, 0.026250),
  rgb(210)=(0.987464, 0.690366, 0.079990),
  rgb(225)=(0.973088, 0.805409, 0.216877),
  rgb(240)=(0.947594, 0.917399, 0.410665),
  rgb(255)=(0.988362, 0.998364, 0.644924),
}}

\definecolor{thsViolet1}{HTML}{E4C7F1} % Purp 2 <https://www.color-name.com/hex/e4c7f1>
\definecolor{thsViolet2}{HTML}{9F82CE} % Purp 5 <https://www.color-name.com/hex/9f82ce>
\definecolor{thsViolet3}{HTML}{4B4B8F} % Bold 11 <https://www.color-name.com/hex/4b4b8f>
\definecolor{thsViolet4}{HTML}{800080} % Purple (HTML/CSS color)  <https://www.color-name.com/hex/800080>

\definecolor{thsPink1}{HTML}{DCB0F2} % Pastel 4 <https://www.color-name.com/hex/dcb0f2>
\definecolor{thsPink2}{HTML}{FE88B1} % Pastel 7 <https://www.color-name.com/hex/fe88b1>
\definecolor{thsPink3}{HTML}{CC3A8E} % Vivid 11 <https://www.color-name.com/hex/cc3a8e>

\definecolor{thsBlue1}{HTML}{4F86F7} % <https://www.color-name.com/blueberry.color>
\definecolor{thsBlue2}{HTML}{0F52BA} % <https://www.color-name.com/sapphire.color>
\definecolor{thsBlue3}{HTML}{332288} % Safe 5 <https://www.color-name.com/hex/332288>
\definecolor{thsBlue4}{HTML}{000080} % Navy <https://www.color-name.com/navy.color>

\definecolor{thsCyan1}{HTML}{88CCEE} % Safe 1 <https://www.color-name.com/hex/88ccee>
\definecolor{thsCyan2}{HTML}{66C5CC} % Pastel 1 <https://www.color-name.com/hex/66c5cc>
\definecolor{thsCyan3}{HTML}{367588} % <https://www.color-name.com/teal-blue.color>

\definecolor{thsGreen0}{HTML}{A7FC00} % <https://www.color-name.com/spring-bud.color>
\definecolor{thsGreen1}{HTML}{4FBF26} % <https://www.color-name.com/perfect-green.color>
\definecolor{thsGreen2}{HTML}{00A86B} % Jade <https://www.color-name.com/jade.color>
\definecolor{thsGreen3}{HTML}{0F8554} % Prism 4 <https://www.color-name.com/hex/0f8554>

\definecolor{thsYellow1}{HTML}{FFFF66} % <https://www.color-name.com/laser-lemon.color>
\definecolor{thsYellow2}{HTML}{FAE500} % <https://www.color-name.com/radioactive.color>
\definecolor{thsYellow3}{HTML}{FFFF00} % <https://www.color-name.com/hex/ffff00>

\definecolor{thsOrange1}{HTML}{ECDA9A} % OrYel 1 <https://www.color-name.com/hex/ecda9a>
\definecolor{thsOrange2}{HTML}{F7945D} % OrYel 3 <https://www.color-name.com/hex/f7945d>
\definecolor{thsOrange3}{HTML}{E17C05} % Prism 7 <https://www.color-name.com/hex/E17C05>
\definecolor{thsOrange4}{HTML}{EC5800} % Persimmon <https://www.color-name.com/persimmon.color>
\definecolor{thsOrange5}{HTML}{F04A00} % Tangelo <https://www.color-name.com/hex/f04a00>

\definecolor{thsRed1}{HTML}{CC503E} % Prism 8 <https://www.color-name.com/hex/cc503e>
\definecolor{thsRed2}{HTML}{FF0000} % Red <https://www.color-name.com/hex/ff0000>
\definecolor{thsRed3}{HTML}{CC0000} % <https://www.color-name.com/hex/CC0000>
\definecolor{thsRed4}{HTML}{670E10} % <https://www.color-name.com/royal-maroon.color>

\definecolor{thsGray1}{HTML}{B1B6B7} % <https://www.color-name.com/antique-steel.color>
\definecolor{thsGray2}{HTML}{828382} % <https://www.color-name.com/neutral-grey.color>
\definecolor{thsGray3}{HTML}{36454F} % <https://www.color-name.com/charcoal.color>

\definecolor{thsBrown1}{HTML}{D6BD8D} % <https://www.color-name.com/hex/d6bd8d>
\definecolor{thsBrown2}{HTML}{BD925A} % <https://www.color-name.com/hex/bd925a>
\definecolor{thsBrown3}{HTML}{A16928} % <https://www.color-name.com/hex/a16928>
\definecolor{thsBrown4}{HTML}{A16928} % <https://www.color-name.com/oak-brown.color>